\documentclass[10pt,twocolumn]{article}

\usepackage[a4paper,top=1.65cm,bottom=1.8cm,left=1.65cm,right=1.65cm,columnsep=0.65cm]{geometry}
\usepackage[T1]{fontenc}
\usepackage[utf8]{inputenc}
\usepackage{libertinus}
\usepackage{microtype}
\usepackage{amsmath,amssymb,mathtools}
\usepackage{graphicx}
\usepackage{booktabs}
\usepackage{multirow}
\usepackage{caption}
\usepackage{subcaption}
\usepackage{dblfloatfix}
\usepackage{placeins}
\usepackage{titlesec}
\usepackage[authoryear,round]{natbib}
\usepackage{newunicodechar}
\usepackage[hidelinks]{hyperref}
\usepackage{fancyhdr}
\usepackage{xcolor}

\usepackage{xparse}

\NewDocumentCommand{\preprintgraphic}{O{}m}{%
  \IfFileExists{#2}{%
    \includegraphics[#1]{#2}%
  }{%
    \fbox{\parbox[c][4.2cm][c]{0.90\linewidth}{%
      \centering\small
      Original figure file \texttt{#2} is not included in this revision package.\\[0.4em]
      Copy it from the existing Overleaf project.
    }}%
  }%
}

\newunicodechar{≤}{\ensuremath{\leq}}
\newunicodechar{≥}{\ensuremath{\geq}}
\newunicodechar{≈}{\ensuremath{\approx}}
\newunicodechar{′}{\ensuremath{^\prime}}

\hypersetup{
  pdftitle={Vertical Attenuation of Thermoremanent Magnetic Anomalies: Implications for Drone-Borne Archaeological and Shallow Geophysical Surveys},
  pdfauthor={Alexandru Hegyi et al.}
}

\titleformat{\section}
  {\large\bfseries\sffamily}{\thesection}{0.55em}{}
\titleformat{\subsection}
  {\normalsize\bfseries\sffamily}{\thesubsection}{0.5em}{}
\titleformat{\subsubsection}
  {\normalsize\itshape}{\thesubsubsection}{0.5em}{}
\titlespacing*{\section}{0pt}{1.6ex plus .4ex}{0.7ex}
\titlespacing*{\subsection}{0pt}{1.3ex plus .3ex}{0.5ex}
\titlespacing*{\subsubsection}{0pt}{1.1ex plus .2ex}{0.4ex}

\begin{document}

% Full-width title and abstract above the two-column article body.
\twocolumn[
\begin{@twocolumnfalse}
\begin{center}
{\fontsize{18}{21}\selectfont\bfseries
Vertical Attenuation of Thermoremanent Magnetic Anomalies:\\
Implications for Drone-Borne Archaeological and Shallow Geophysical Surveys\par}
\vspace{0.75em}

{\normalsize
Alexandru Hegyi\textsuperscript{1,2,*},
Kristoffer Aalstad\textsuperscript{1},
Arne Anderson Stamnes\textsuperscript{3},
Luc Girod\textsuperscript{1},
Sebastian Westermann\textsuperscript{1},\\
Diaa Sheishah\textsuperscript{4,5,*},
Enas Abdelsamei\textsuperscript{4,5},
Michał Pisz\textsuperscript{6},
and Norbert Pirk\textsuperscript{1}\par}
\vspace{0.7em}

\begin{minipage}{0.94\textwidth}
\small
\textsuperscript{1}Department of Geosciences, University of Oslo, Sem Sælands vei 1, 0371 Oslo, Norway\\
\textsuperscript{2}Institute for Advanced Environmental Research (ICAM), West University of Timișoara, Oituz 4c, 300086 Timișoara, Romania\\
\textsuperscript{3}The Norwegian Institute for Cultural Heritage Research (NIKU), Storgata 2, 0105 Oslo, Norway\\
\textsuperscript{4}University of Szeged, Department of Physical and Environmental Geography, 6722 Szeged, Egyetem u. 2--6, Hungary\\
\textsuperscript{5}National Research Institute of Astronomy and Geophysics, El Marsad St., Helwan, Cairo 11421, Egypt\\
\textsuperscript{6}School of Archaeological and Forensic Sciences, University of Bradford, Bradford BD7 1DP, UK\\
\vspace{0.35em}
\textsuperscript{*}\textit{Corresponding authors:}
\href{mailto:alexandru.hegyi@geo.uio.no}{alexandru.hegyi@geo.uio.no};
\href{mailto:geo_diaa@nriag.sci.eg}{geo\_diaa@nriag.sci.eg}
\end{minipage}
\end{center}

\vspace{0.65em}
\hrule
\vspace{0.65em}

\begin{minipage}{0.96\textwidth}
\small
\noindent\textbf{Abstract.} Magnetic surveys are widely used in archaeology to reveal buried features by detecting variations in the Earth’s magnetic field caused by past human activities. However, as sensor height increases, magnetic anomalies produced by subsurface sources weaken and broaden, posing challenges for both ground-based and aerial (UAV) surveys. This study quantifies that altitude-dependent loss of signal and spatial definition using a fully synthetic model of a uniformly magnetized circular kiln together with upward continuation of measured fluxgate data from the Storbekken~1 iron-production site in central Norway. The synthetic experiment shows that a 2\,m diameter, 1\,m thick magnetized cylinder buried 0.3\,m below ground retains only about 11\% of its near-surface (0.2\,m) peak amplitude at 2\,m sensor height and about 3\% at 4\,m. In the hybrid experiment, composite furnace anomalies of several hundred nanoteslas collapse to tens of nanoteslas by 2--3\,m, while distinct lobes merge into broad, low-contrast features. The results show that UAV-borne magnetometry can remain effective for site detection and delineation at elevations where detailed intra-site morphology is already strongly degraded. For small or subtle archaeological targets, sub-metre to low-metre sensor heights provide substantially greater interpretative detail. Recent field evidence on UAV-system noise further indicates that practical detectability will depend not only on source-distance attenuation but also on platform configuration and acquisition strategy. The resulting attenuation curves and morphological observations provide transferable guidance for planning archaeological and shallow-geophysical UAV magnetic surveys.

\vspace{0.45em}
\end{minipage}

\vspace{0.9em}
\hrule
\vspace{1.0em}
\end{@twocolumnfalse}
]

\section{Introduction}

Magnetic prospection has become a cornerstone of archaeological surveying, particularly for identifying and characterizing ancient iron production sites \citep{kvamme_magnetometry_2006, johnson_chris_2004, fassbinder_seeing_2015}. Iron smelting processes, such as bloomery and blast furnace operations, impart distinct thermoremanent magnetization (TRM) signatures to furnace linings, slags, and surrounding heated soils. These magnetic anomalies are typically strong, spatially discrete, and can provide critical insights into subsurface archaeological features without invasive excavation \citep{Powell2008, Abrahamsen2002, Crew2002}.

Prehistoric bloomery furnaces, which operated at temperatures around \(1100\text{–}1400\,^\circ\) C, generally generate present day prominent magnetic anomalies in the range of \(+500\text{–}+2000\,\mathrm{nT}\) measured near the surface. These anomalies are typically accompanied by flanking negative values in adjacent thermally altered soils and slags \citep{Abrahamsen2002, Crew2002, stamnes_magnetic_2019}. In contrast, medieval and post-medieval blast furnaces operated at temperatures approaching \(1500\,^\circ\)C and produced considerably larger slag volumes, generating significantly stronger magnetic anomalies. Typical magnetic signatures at these sites range between \(+2000\) and \(+5000\,\mathrm{nT}\), extending over more extensive spatial footprints \citep{Vernon1999, Walach2011, Kozhevnikov2018}.

Understanding how these magnetic anomalies attenuate with increasing sensor altitude is crucial in the context of unoccupied aerial vehicle (UAV) based magnetic surveys. UAV magnetometry has recently become a highly effective method for rapidly surveying extensive archaeological landscapes at sensor altitudes commonly between \(2\) and \(6\,\mathrm{m}\) above ground. However, a major challenge stems from the rapid decay of magnetic anomaly amplitudes with increasing sensor height. For example, a bloomery anomaly of \(+1000\,\mathrm{nT}\) at \(0.5\,\mathrm{m}\) depth can fall below \(+300\,\mathrm{nT}\) when surveyed at a height of \(3\,\mathrm{m}\) above ground level (AGL), potentially leading to missed features or misclassifications if measurements remain uncorrected for altitude \citep{Powell2008}. Controlled test flights have confirmed that anomalies can still be detected several meters above ground, but they exhibit significant lateral broadening and reduced peak intensity, necessitating careful flight‐path planning based on high‐resolution surface models \citep{Schmidt2024}. Similarly, recent drone-borne surveys have revealed that optimal detection requires flying as low as \(2\text{–}3\,\mathrm{m}\) and applying comprehensive noise‐filtering post-processing procedures \citep{Stele2023}.

Further advancements in UAV magnetometry have addressed specific technical challenges. Vertical gradiometer implementations, where the sensor is suspended \(3\,\mathrm{m}\) below a UAV, have highlighted issues of oscillation induced noise and magnetic interference from the platform. Researchers have successfully mitigated these by using Continuous Wavelet Transform (CWT) analysis and designing low‐pass filters tailored to dominant oscillation frequencies, allowing for the successful mapping of subtle archaeological features \citep{DiGiacomo2024}. Moreover, integrated ground and drone survey workflows have proven effective in improving feature classification accuracy. Combining high-resolution ground-based gradiometry with complementary drone-borne total field and gradiometric data, particularly when employing machine learning driven processing, has enhanced the separation of overlapping anomaly sources in complex archaeological sites \citep{Gavazzi2021, Rossi2023}.

Building on these insights from recent test cases, the present study first generates a purely synthetic magnetic anomaly model, representing a uniformly magnetized circular cylinder (similar to a kiln). This model is used to quantitatively theoretically assess magnetic anomaly decay and lateral widening as sensor height increases. Once evaluated, this model is then applied to real magnetic raster data, acquired at ground level (0.2 m sensor elevation) over an Iron Age production site in central Norway located at 0.3 m depth \citep{stamnes_magnetic_2019}. By applying frequency‐domain upward continuation and adding Gaussian noise consistent with modern UAV‐borne gradiometers (\(0.004\,\mathrm{nT}/\sqrt{\mathrm{Hz}}\) at \(10\,\mathrm{Hz}\)), we simulate synthetic "drone" surveys at various sensor heights ranging from \(0.2\) m to \(15.0\) m. Extracting lateral profiles and peak‐amplitude versus height curves in these observing system simulation experiments allows for precise quantification of anomaly attenuation and spatial smearing at typical UAV altitudes.

This combined synthetic and empirical approach grounded in recent field tests \citep{Stele2023, Schmidt2024, DiGiacomo2024, Gavazzi2021, Rossi2023}, provides practical guidelines for designing experiments involving UAV survey planning, including optimal flight altitude selection, instrument settings, and post‐processing workflows. Ultimately, this study aims to enhance the efficiency, precision, and interpretative power of drone‐borne magnetic prospections in complex iron‐production landscapes and, by analogy, any thermally altered archaeological or shallow subsurface geophysical features.

Despite recent advances in UAV-borne magnetometry, a critical research gap remains in quantitatively understanding how archaeological magnetic anomalies deteriorate with increasing sensor altitude. Previous studies have demonstrated the feasibility of drone-based surveys and addressed key technical challenges such as platform-induced noise, sensor oscillation, and filtering strategies. However, these investigations are largely case-specific and do not provide a systematic, quantitative assessment of anomaly amplitude decay and morphological broadening as a function of flight height, nor do they establish altitude-dependent detectability thresholds under realistic noise conditions. Consequently, survey design and altitude selection remain largely empirical.

To address this gap, the present study introduces a novel integrative framework that combines controlled synthetic modeling of thermoremanent archaeological sources with upward continuation of real magnetic field data. This approach enables precise quantification of both amplitude attenuation and spatial smoothing with increasing sensor height, while explicitly accounting for realistic UAV noise levels. By doing so, the study establishes quantitative decay curves and morphological guidelines that define the practical limits of drone-borne magnetometry for site detection versus intra-site feature characterization. These results extend the current state of the art by transforming qualitative observations into transferable, evidence-based criteria for UAV survey design in archaeological and shallow geophysical applications.

\section{Theoretical Framework}

Magnetic anomalies over archaeological kilns and hearths arise primarily from thermoremanent magnetization acquired by heated materials (slag, baked clay) during past metallurgical activities \citep{fassbinder_seeing_2015}. To interpret and predict how such anomalies attenuate with sensor height, we draw upon classical potential field theory \citep{armitage_classical_2012}, spectral continuation methods, and noise characteristics of fluxgate magnetometers. Below, we summarize the key theoretical elements underpinning both our synthetic and hybrid workflows.

\subsection{Dipole Approximation of a Kiln Anomaly}
An isolated kiln or hearth of limited lateral extent can be approximated, to a first order, as a magnetic dipole embedded within a homogeneous, non‐magnetic half‐space. If the dipole moment vector is \(\mathbf{m}=(m_x,\,m_y,\,m_z)\) located at depth \(D\) below the surface, the magnetic induction at an observation point \(\mathbf{r}=(x,\,y,\,z)\) is given by Eq. ~\ref{eq:dipole_field}.
\begin{equation}\label{eq:dipole_field}
\begin{aligned}
\mathbf{B}(\mathbf{r})
&= \frac{\mu_{0}}{4\pi}
\Biggl[
\frac{3(\mathbf{m}\cdot\Delta\mathbf{r})\Delta\mathbf{r}}
{\lvert\Delta\mathbf{r}\rvert^{5}} \\
&\qquad - \frac{\mathbf{m}}{\lvert\Delta\mathbf{r}\rvert^{3}}
\Biggr], \\
\Delta\mathbf{r} &= \mathbf{r}-\mathbf{r}_0.
\end{aligned}
\end{equation}
where \(\mu_{0}\) is the permeability of free space and $\mathbf{r}_0=(x_{0},\,y_{0},z_0)$ is the dipole's position vector with a dipole elevation $z_0=-D$ and dipole horizontal location \((x_{0},\,y_{0})\). In an archaeological survey, we typically measure only the vertical component \(B_{z}\) even if total field measurements also render excellent results \citep{fassbinder_seeing_2015}. For a dipole oriented vertically \(\mathbf{m} = m_z\,\hat{\mathbf{z}}\) (i.e., $m_x=m_y=0$) at depth \(D\) (i.e. elevation $z_0=-D$) and horizontal location $x_0=y_0=0$, the field at height \(h\) above ground \((z =h)\) at horizontal position \(x=y=0\) simplifies to (Eq. ~\ref{eq:Bz_peak}):

\begin{equation}\label{eq:Bz_peak}
  B_{z}(0,0,h)
  = \frac{\mu_{0}}{4\pi}\,m_z\,\frac{2}{(h + D)^{3}}
  \;\propto\;(h + D)^{-3}\,.
\end{equation}
since $\Delta \mathbf{r}=(h+D)\hat{\mathbf{z}}$ in this case. Thus, peak anomaly amplitude decays approximately as the inverse cube of the total distance \((h + D)\). Deviations from ideal dipole behavior in real data (heterogeneous magnetization, finite source geometry) introduce second-order effects, but the \(\sim1/(h+D)^{3}\) law remains a robust first-order approximation \citep{bevan_magnetic_1994, blakely_potential_1996, langouet_clark_1991}.

\subsection{Upward Continuation in the Spectral Domain}
Upward continuation predicts the magnetic field at a higher elevation from measurements at a lower plane under the assumption of a homogeneous, current‐free half‐space. If \(B_{z}(x,y,0)\) is the vertical component on the ground plane, its two‐dimensional Fourier transform (Eq.~\ref{eq:fourier_Bz}) is
\begin{equation}\label{eq:fourier_Bz}
\begin{aligned}
\widetilde{B}(k_x,k_y;0)
 &= \iint_{-\infty}^{\infty} B_z(x,y,0) \\
 &\quad \times \exp\!\bigl[-\mathrm{i}(k_xx+k_yy)\bigr]\,
 \mathrm{d}x\,\mathrm{d}y.
\end{aligned}
\end{equation}
where \((k_{x},\,k_{y})\) are horizontal wavenumbers. At elevation \(z = h\), potential‐field theory dictates
\begin{equation}\label{eq:upward_continuation}
\begin{aligned}
\widetilde{B}(k_x,k_y;h)
 &= \widetilde{B}(k_x,k_y;0)\,e^{-Kh},\\
K &= \sqrt{k_x^2+k_y^2}.
\end{aligned}
\end{equation}
Equation~\ref{eq:upward_continuation} shows that each horizontal wavenumber component decays as \(\exp\!\bigl(-K\,h\bigr)\), so higher frequency (short wavelength) features are increasingly suppressed with height.
Consequently, each spectral component decays exponentially with height \(\exp\!\bigl(-K\,h\bigr)\). In practice, given a discrete grid \(B_{z}[i,j]\) of size \(N_{y}\times N_{x}\) and cell spacing \(\Delta x\), one implements:

\begin{enumerate}
  \item Compute the 2D FFT, \(G[u,v] = \mathrm{FFT}\bigl(B_{z}\bigr)\), with associated wavenumbers \(k_{x}(u),\,k_{y}(v)\) determined by \(\mathrm{fftfreq}(N,\Delta x)\).
  \item Form the filter \(H[u,v] = \exp\bigl(-\sqrt{k_{x}(u)^{2} + k_{y}(v)^{2}}\;h\bigr)\).
  \item Multiply \(G_{\mathrm{cont}}[u,v] = G[u,v]\;H[u,v]\).
  \item Apply the inverse FFT to obtain \(B_{z}(x,y;h) = \Re\{\mathrm{IFFT}(G_{\mathrm{cont}})\}\).
\end{enumerate}
Masked or no‐data cells are temporarily set to zero to minimize spectral leakage; the original mask is reapplied after continuation. By suppressing high wavenumber (short wavelength) components, upward continuation naturally smooths the anomaly map, preserving long wavelength trends while attenuating fine detail as \(h\) increases \citep{bhattacharyya_general_1967, phillips_potentialfield_1996, nabighian_historical_2005}.

\subsection{Fluxgate Noise Modeling}
Fluxgate gradiometers exhibit noise spectral densities on the order of \(1\text{–}10\,\mathrm{pT}/\sqrt{\mathrm{Hz}}\). When sampling at frequency \(f_{s}\), the one-sided Nyquist bandwidth is \(f_{s}/2\), so the per-sample standard deviation of white noise (Eq.~\ref{eq:white_noise}) is
\begin{equation}\label{eq:white_noise}
  \sigma_{\mathrm{white}}
  = N_{d}\,\sqrt{\frac{f_{s}}{2}}.
\end{equation}
where \(N_{d}\) is the sensor noise density in \(\mathrm{nT}/\sqrt{\mathrm{Hz}}\). In our simulations, we use \(N_{d} = 0.004\,\mathrm{nT}/\sqrt{\mathrm{Hz}}\) and \(f_{s} = 10\,\mathrm{Hz}\). As shown in Eq.~\ref{eq:white_noise_value}, the per-sample noise standard deviation is approximately 0.00894 nT.
\begin{equation}\label{eq:white_noise_value}
  \sigma_{\mathrm{white}}
  \approx 0.004\,\sqrt{\frac{10}{2}}
  = 0.00894\;\mathrm{nT}.
\end{equation}
Real fluxgate noise exhibits spatial and temporal correlation due to finite coil aperture and onboard filtering. We approximate this by convolving the white noise field with a two-dimensional Gaussian kernel of standard deviation \(\sigma_{\mathrm{spatial}}=5\) pixels. The resulting spatially correlated noise map mimics typical aperture-induced smoothing observed in UAV‐borne systems \citep{ocker_magnetometer_2014, sunderland_characterising_2009, ciminale_aspects_2001}.

\subsection{Altitude-Dependent Anomaly Attenuation}
The central objective of the synthetic experiment is to quantify how anomaly amplitude and morphology change as a function of sensor height \(h\). The source is a uniformly magnetized circular cylinder (diameter \(2.0\,\mathrm{m}\), thickness \(1.0\,\mathrm{m}\)) buried with its top face at \(0.30\,\mathrm{m}\) depth. For each height, we compute the upward-continued field \(B_z(x,y;h)\), including the imposed soil trend, spatial heterogeneity, and instrument-noise realization, and record the global absolute peak:
\begin{equation}\label{eq:peak_extraction}
  P_{\mathrm{syn}}(h)
  = \max_{x,y}\bigl\lvert B_{z}(x,y;h)\bigr\rvert .
\end{equation}
Attenuation is reported directly relative to the lowest simulated observation height \(h_{\mathrm{ref}}=0.2\,\mathrm{m}\):
\begin{equation}\label{eq:synthetic_retention}
  R_{\mathrm{syn}}(h)
  = 100\,\frac{P_{\mathrm{syn}}(h)}{P_{\mathrm{syn}}(h_{\mathrm{ref}})} .
\end{equation}
No parametric decay law is fitted to the numerical experiments. The inverse-cube relationship in Eq.~\ref{eq:Bz_peak} is retained only as the analytical far-field reference for an ideal point dipole. The finite synthetic source, imposed background structure, and real multi-source field data are instead evaluated through their directly simulated or continued amplitudes and spatial morphology.

\section{Materials and Methods}

\subsection{Computational Implementation and Software Framework}
All numerical simulations, spectral upward continuations, noise realizations, attenuation measurements, and visualizations were implemented using custom Python scripts developed for this study and released as open-source software in the MagSim repository. The synthetic workflow (\textit{KilnSim.py}) forward-models the magnetic response of a buried, uniformly magnetized kiln represented by an array of dipoles, superimposed with soil trends, spatial heterogeneity, and fluxgate-equivalent noise, before applying FFT-based upward continuation and direct peak-amplitude extraction. The hybrid workflow (\textit{AirSim\_Storbekken1\_Norway.py}) operates directly on measured magnetic-gradient rasters, extracts user-defined transects, simulates increased sensor height via spectral continuation, injects the specified noise realization, and quantifies height-dependent attenuation along profile corridors. The analysis reported here uses the directly simulated or continued amplitudes and normalized retention values rather than a parametric decay fit.

\subsection{Synthetic Data Generation}

\subsubsection{Kiln model and discretization}  
\begin{figure*}[!t]
\centering
\preprintgraphic[width=0.96\textwidth]{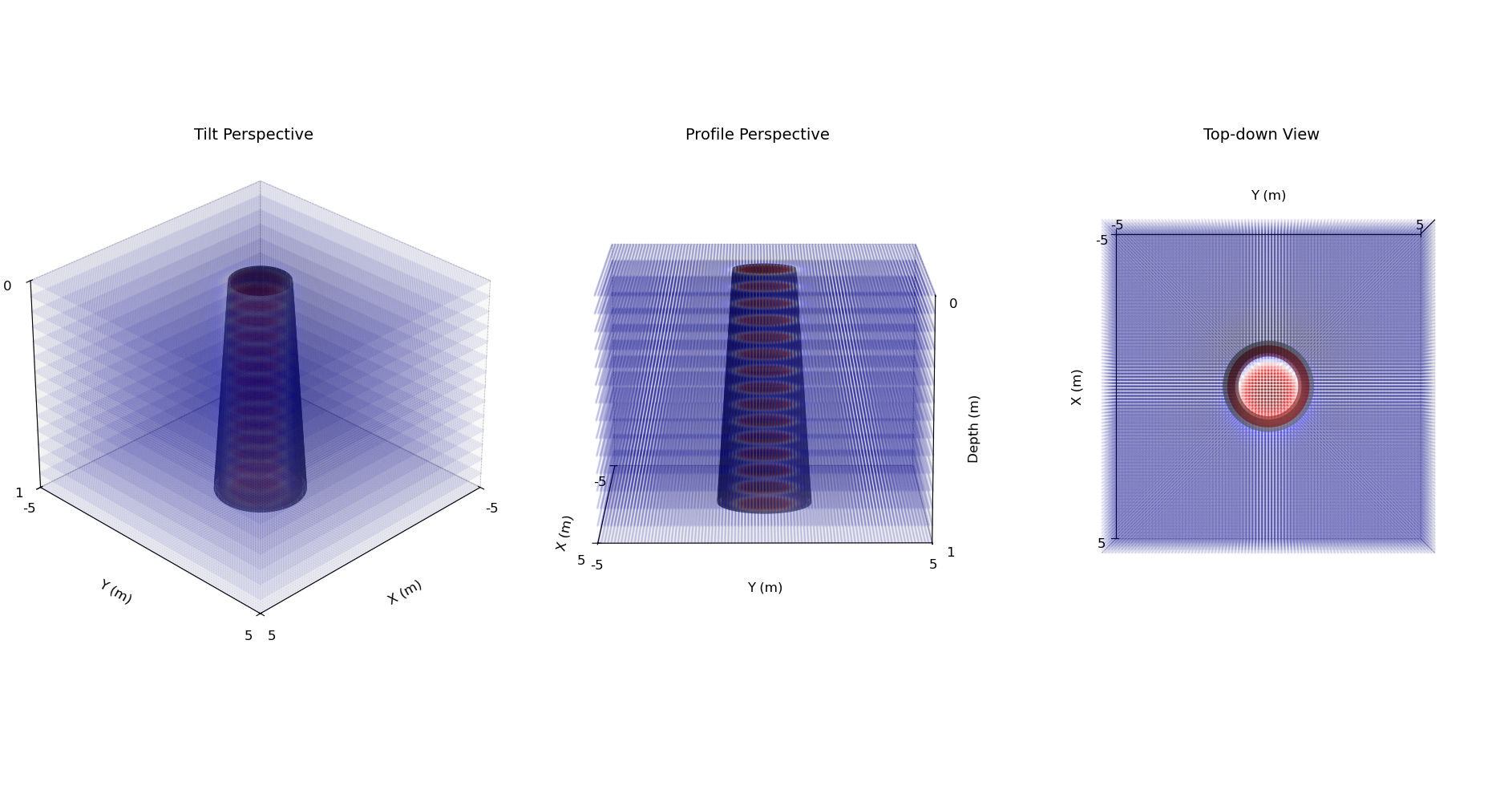}
\caption{Three-dimensional visualization of the synthetic kiln model and its associated magnetic anomaly ($B_z$) volume. The panels show a tilted perspective (left), profile perspective (centre), and top-down view (right). Dark grey and red surfaces represent the kiln structure, while the semi-transparent scatter layers depict the $B_z$ anomaly in the surrounding volume.}
\label{fig:synthetic_kiln_3d_model}
\end{figure*}

We represented a buried kiln as a uniformly magnetized circular cylinder (Fig ~\ref {fig:synthetic_kiln_3d_model}) of diameter \(D = 2.0\)\,m and thickness \(T = 1.0\)\,m, whose top lies at depth \(d_{\mathrm{top}} = 0.30\)\,m. The survey domain is a square of side length \(L = 10.0\)\,m discretized at \(\Delta x = 0.10\)\,m (yielding 101$\times$101 grid points). All grid cells center coordinates \((x,y)\) satisfying \((x-x_0)^{2}+(y-y_{0})^{2}\le (D/2)^{2}\) were assigned a magnetic sub-dipole of moment (Eq.~\ref{eq:msub})
\begin{equation}\label{eq:msub}
\begin{aligned}
m_{\mathrm{sub}}
 &= M(\Delta x)^2T \\
 &= 1.5\,\mathrm{A/m}\,(0.10\,\mathrm{m})^2(1.0\,\mathrm{m}) \\
 &= 0.015\,\mathrm{A\,m^2}.
\end{aligned}
\end{equation}
aligned with the ambient geomagnetic field (inclination \(I=70^\circ\), declination \(D=3^\circ\)) and situated at depth \(d_{\mathrm{sub}}=d_{\mathrm{top}}+T/2=0.80\)\,m. The vertical component \(B_{z}(x,y;0)\) was computed via the standard dipole kernel (\(\mu_{0}/4\pi=10^{-7}\), SI), then converted to nanotesla \citep{bhattacharyya_general_1967,phillips_potentialfield_1996}.

\subsubsection{Addition of soil trend, heterogeneity, and instrument noise}  
To \(B_{z}(x,y;0)\) we superimposed:
\begin{enumerate}
  \item A uniform linear trend of \(+1.0\,\mathrm{nT/m}\) along the \(x\) axis.
  \item Spatial heterogeneity: Gaussian noise with standard deviation \(\sigma_{\mathrm{het}}=0.5\)\,nT, smoothed by a Gaussian kernel (\(\sigma_{\mathrm{space}}=10\) pixels) to mimic soil variability \citep{ocker_magnetometer_2014}.
\item Instrumental noise: we first generate Gaussian noise in time with spectral density $N_d=0.004\,$nT/$\sqrt{\mathrm{Hz}}$ at $f_s=10\,$Hz.
We then spatially smooth that time series with a Gaussian filter (width $\sigma_{\mathrm{corr}}=5$ pixels) to approximate the fluxgate’s spatial correlations, turning the originally “white” (uncorrelated) noise into “colored” (spatially correlated) noise \citep{sunderland_characterising_2009,ciminale_aspects_2001}.
\end{enumerate}

\subsubsection{Spectral upward continuation}

To simulate measurements at altitude \(h\), we applied the spectral filter (Eq.~\ref{eq:upward_filter})
\begin{equation}\label{eq:upward_filter}
  \tilde{B}_{z}(k_{x},k_{y};h)
  = \tilde{B}_{z}(k_{x},k_{y};0)\,\exp\bigl(-\sqrt{k_{x}^{2}+k_{y}^{2}}\,h\bigr).
\end{equation}
Here \(\tilde{B}_{z}\) denotes the 2D Fourier transform of \(B_{z}\) on the \(N_{y}\times N_{x}\) grid with spacing \(\Delta x\). Taking the inverse FFT then yields \(B_{z}(x,y;h)\). The continuation heights were set to (Eq.~\ref{eq:continuation_heights})
\begin{equation}\label{eq:continuation_heights}
\begin{aligned}
h=\{&0.2,0.3,0.4,0.5,1.0,1.5,2.0,2.5,\\
     &3.0,3.5,4.0,4.5,5.0,6.0,10.0,15.0\}\,\mathrm{m}.
\end{aligned}
\end{equation}

After each continuation, a fresh realization of instrument noise, smoothed by 5 pixels, was added to approximate the fluxgate’s spatial correlations.

\subsubsection{Global peak extraction and relative retention}

For each height \(h\), the global anomaly amplitude was computed as
\begin{equation}\label{eq:Psyn}
  P_{\mathrm{syn}}(h)
  = \max_{x,y}\bigl\lvert B_{z}(x,y;h)\bigr\rvert .
\end{equation}
The values were then normalized to the \(h=0.2\,\mathrm{m}\) reference height using Eq.~\ref{eq:synthetic_retention}. This direct representation avoids imposing a single parametric decay law on a finite source embedded in a spatially variable background. At the greatest continuation heights the source contribution becomes sufficiently weak that the global maximum can increasingly reflect the imposed background field and residual noise; those values therefore represent a practical background-dominated level rather than a pure kiln-source amplitude.

\subsection{Field Data and Hybrid Continuation}
\subsubsection{Site Description: Storbekken 1}

The following description of Storbekken 1 is based primarily on \citet{stamnes_magnetic_2019}. Storbekken 1 lies on a gently sloping terrace above a tributary of the Gaula River at Tovmoen, Budalen, Midtre Gauldal, Sør-Trøndelag, central Norway (62°56′ N, 10°17′ E; 365 m a.s.l.), in mixed conifer woodlands over glacial till. In 1988, visual inspection and auger probing revealed five stone-lined slag-pit furnaces (Furnaces I–V), each 1.2–1.5 m in diameter, aligned NE–SW at ~5.0–5.5 m spacing; fan-shaped slag tips extend downslope. A 1993 excavation of Furnace II exposed two intact pits containing ~71 kg of in situ slag; charcoal from the furnace floor yielded a calibrated radiocarbon date of BC 180–AD 25 (2050 ± 85 BP). The other furnaces are visible as oblong shallow depressions and working pits are visible as smaller, round and shallow pits surrounding the furnace \citep{stamnes_magnetic_2019}.

\begin{figure*}[!t]
\centering
\preprintgraphic[width=0.96\textwidth]{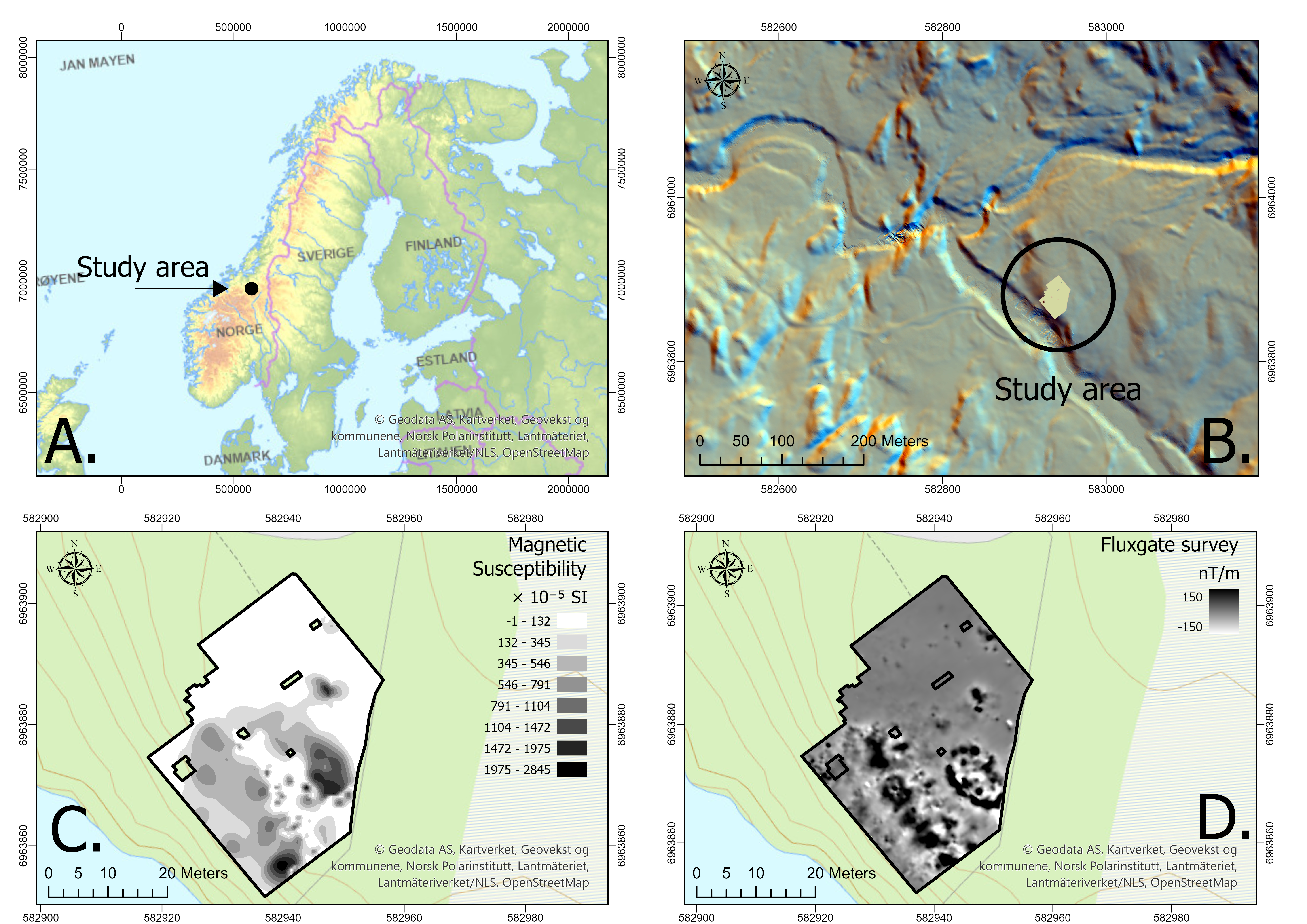}
\caption[Geophysical survey results at Storbekken~1]{Geophysical survey results at the Storbekken~1 site, reprocessed using the raw data published by \citet{stamnes_magnetic_2019}. (A) Regional location of the study area in Norway on a shaded-relief map of Scandinavia. (B) Local topographic context shown with multi-illumination hillshade. (C) High-resolution magnetic-susceptibility map; the black polygon delineates the fluxgate-gradiometer survey. (D) Fluxgate-gradiometer map (grayscale, $\pm150$ nT/m) displayed over the topographic background.}
\label{fig:storbekken1_surveys}
\end{figure*}

In autumn 2014, topsoil volume susceptibility (MS) was measured with a Bartington MS2 D-loop over 7570 m² at 3.44 m spacing (Figure~\ref{fig:storbekken1_surveys}). Excavated Furnace II recorded MS = 1184–1833 \(\times10^{-5}\) SI (mean 1508.5 \(\times10^{-5}\) SI); unexcavated furnaces (I, III–V) ranged 339–1560 \(\times10^{-5}\) SI (mean 929.3 \(\times10^{-5}\) SI). The main slag tip measured 12–3226 \(\times10^{-5}\) SI (mean 452.8 \(\times10^{-5}\) SI), ≈ 43× local background (10.5 \(\times10^{-5}\) SI). An earth bank (interpreted as roasted iron ore storage) registered 824–2587 \(\times10^{-5}\) SI (mean 1618 \(\times10^{-5}\) SI), exceeding even the unexcavated furnaces. Two previously undetected hotspots (A1, A2; each ~5 m × 6 m) lay north of the furnaces: A1 averaged 803.2 \(\times10^{-5}\) SI (range 28–2014), A2 averaged 150.7 \(\times10^{-5}\) SI (range 13–453), with no surface trace \citep{stamnes_magnetic_2019}.

Simultaneously, a fluxgate gradiometer survey (autumn 2014) covered 1221 m² with 0.5 m traverses and 0.125 m sample spacing, recording vertical-gradient anomalies (Fig.~\ref{fig:storbekken1_surveys}). Unexcavated furnaces produced positive peaks of +277 to +318 nT and negative halos of –139 to –128 nT; the excavated Furnace II still exhibited a smaller positive anomaly. Fan-shaped slag heaps generated +260 to +300 nT peaks with –210 nT downslope halos. The earth bank (12 m × 7.5 m) yielded +555 nT/–277 nT anomalies, consistent with roasted-ore storage. Hotspots A1 and A2 corresponded to compact highs of +363 nT/–87 nT and +320 nT/–62 nT (each ~3 m × 2 m), likely roasting sites. These are usually flat surfaces where the roasting activity occurred, and are not necessarily associated with buried features as such  \citep{stamnes_magnetic_2019}.

Combined MS and FG data indicate that the core kiln complex (five furnaces plus pits) spans ~1940 m². MS anomalies extend ~30 m upslope from the furnace row, suggesting secondary roasting or working areas. MS values >2000 \(\times10^{-5}\) SI and FG peaks >+300 nT reliably pinpoint furnace locations; lower anomalies (7–27× background) mark broader activity. Secondary hotspots A1 and A2 (each ~30 m²) lie ~5 m north of the furnaces and required ~3.9 m sampling density to detect \citep{stamnes_magnetic_2019}.

\subsubsection{Transect sampling and upward continuation}

Two linear transects were defined through reference point $(x_0,y_0)$: Profile 1 with azimuth $\theta_1=50^\circ$ and perpendicular offset $\delta_1=-3.5$ m from $(x_0,y_0)$, and Profile 2 with $\theta_2=140^\circ$ and $\delta_2=-20.0$ m. A $\pm2$ m corridor around each transect was selected by including all pixels whose perpendicular distance to the line satisfied:
\begin{equation}
\label{eq:corridor}
\begin{aligned}
d(x,y) &= \bigl|(x-x_0)\sin\theta-(y-y_0)\cos\theta \\
       &\qquad -\delta\bigr| \leq 2\,\mathrm{m}.
\end{aligned}
\end{equation}

where $\theta$ is the azimuth angle measured clockwise from the positive $x$-axis (east), and $\delta$ is the signed perpendicular offset of the line from the reference point $(x_0,y_0)$.

For each height \(h\in\{0.2,0.4,\dots,15.0\}\)\,m, the masked data were upward‐continued by \(\Delta h=h-0.2\)\,m via the FFT filter (Eq.~\ref{eq:upward_filter}) at the prescribed heights (Eq.~\ref{eq:continuation_heights}), then white‐noise (\(\sigma_{\mathrm{white}}\)) smoothed over five pixels was added. Within the corridor \(\mathcal{C}\), the minimum and maximum anomalies were recorded:

Within the corridor $\mathcal{C}$, both the minimum and maximum anomaly amplitudes were tracked as functions of height:
\begin{equation}
\label{eq:peak_max}
P_{\mathrm{obs}}^{\mathrm{max}}(h) = \max_{i\in\mathcal{C}} B_z(i;h),
\end{equation}

\begin{equation}
\label{eq:peak_min}
P_{\mathrm{obs}}^{\mathrm{min}}(h) = \min_{i\in\mathcal{C}} B_z(i;h),
\end{equation}
in direct analogy to the global peak extraction of Eq.~\eqref{eq:Psyn}.

For each profile and each continuation height, the positive and negative extrema were compared directly with their values at the 0.2\,m reference height. Relative attenuation is therefore reported as a percentage of the corresponding baseline amplitude, without fitting a parametric decay model.

\section{Results}

\subsection{Synthetic Anomaly Continuation Experiment}
Our synthetic dipole model, designed to simulate a buried furnace, reveals the characteristic behavior of a compact, circular magnetic anomaly ($B_z$) as it's upward-continued.

\subsubsection{Plan-view Maps and Amplitude Decay}
\begin{figure*}[!t]
\centering
\preprintgraphic[width=0.96\textwidth]{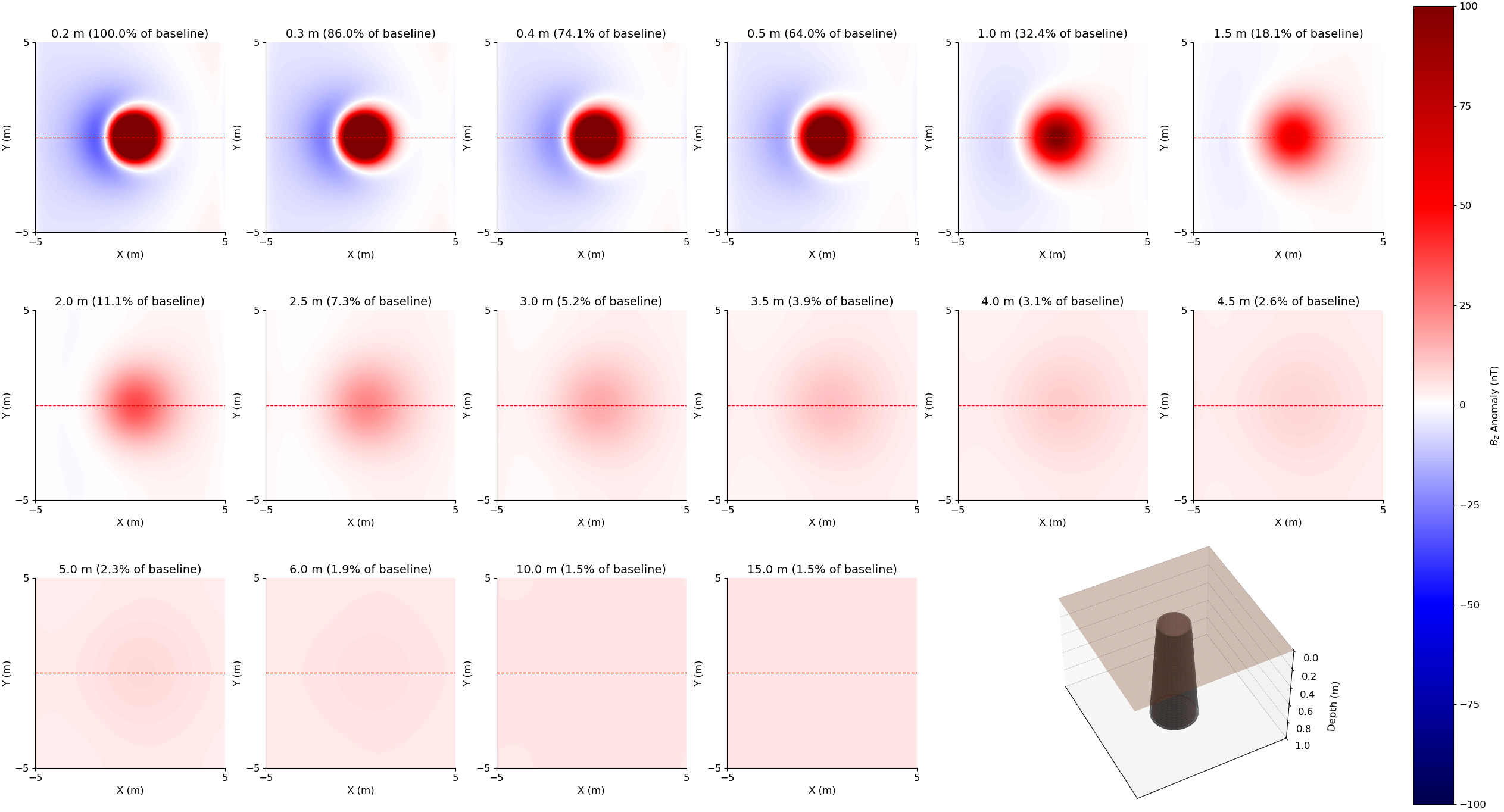}
\caption{Upward-continued $B_z$ anomaly maps at increasing sensor heights from $h=0.2$\,m to $h=15.0$\,m. Each panel shows the anomaly over the synthetic survey grid and illustrates progressive amplitude attenuation and lateral broadening with elevation.}
\label{fig:synthetic_maps}
\end{figure*}

At a near-surface height of $h=0.2$\,m (as shown in Figure~\ref{fig:synthetic_maps}), the synthetic anomaly exhibits a peak amplitude of 322.7\,nT (100\% of baseline), characterized by a full-width at half-maximum (FWHM) of approximately 1.5\,m and a classic dipolar fall-off. As the simulated sensor height increases, both the amplitude and spatial characteristics of the anomaly undergo significant changes.

Quantitative observations of this decay are summarized in Table~\ref{tab:synthetic_decay}. At $h=0.5$\,m, the peak amplitude declines to 206.4\,nT, representing a 36\% reduction. This rapid attenuation continues with increasing height: by $h=1.0$\,m, the amplitude is further reduced to 104.5\,nT (32.4\%), and at $h=2.0$\,m (also visible in Figure~\ref{fig:synthetic_maps}), it reaches only 35.8\,nT (11.1\%) of its original value. At $h=5.0$\,m, the anomaly is merely a faint dome of 7.3\,nT (2.2\%). Notably, the extracted global peak approaches 4.8\,nT at both 10 and 15\,m. At these continuation heights the compact kiln contribution has become very weak, so the global maximum increasingly reflects the imposed broad background trend together with residual spatial heterogeneity and instrument noise. The high-altitude plateau should therefore be interpreted as a background-dominated level in this synthetic experiment rather than as the remaining amplitude of the kiln anomaly alone.
\begin{table}[!t]
\centering
\caption{Global peak amplitude of the synthetic anomaly as a function of continuation height. Percentages are relative to the $h=0.2$\,m baseline.}
\label{tab:synthetic_decay}
\small
\setlength{\tabcolsep}{5pt}
\begin{tabular}{@{}rrr@{}}
\toprule
Height (m) & Peak (nT) & Retention (\%) \\
\midrule
0.20 & 322.7 & 100.0 \\
0.30 & 277.5 & 86.0 \\
0.40 & 239.0 & 74.1 \\
0.50 & 206.4 & 64.0 \\
1.00 & 104.5 & 32.4 \\
1.50 & 58.4 & 18.1 \\
2.00 & 35.8 & 11.1 \\
2.50 & 23.7 & 7.3 \\
3.00 & 16.8 & 5.2 \\
3.50 & 12.7 & 3.9 \\
4.00 & 10.1 & 3.1 \\
4.50 & 8.4 & 2.6 \\
5.00 & 7.3 & 2.2 \\
6.00 & 6.0 & 1.9 \\
10.00 & 4.8 & 1.5 \\
15.00 & 4.8 & 1.5 \\
\bottomrule
\end{tabular}
\end{table}

\subsubsection{Profile Evolution and Direct Attenuation}
\begin{figure}[!t]
\centering
\preprintgraphic[width=\columnwidth]{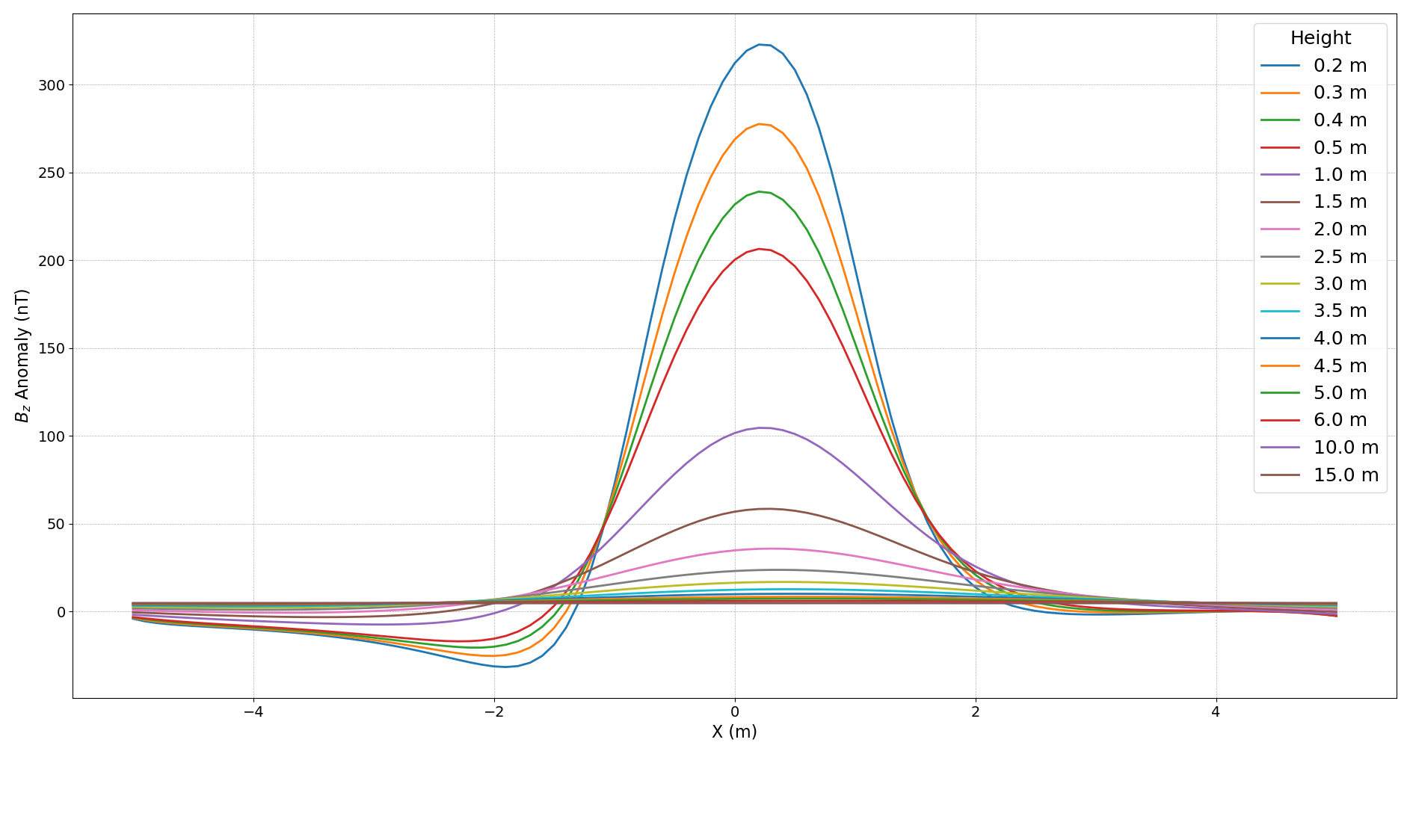}
\caption{Synthetic $B_z$ profiles along $y=0$ for selected continuation heights, showing rapid amplitude attenuation and progressive lateral broadening.}
\label{fig:synthetic_profiles}
\end{figure}

Horizontal profiles along $y=0$ (Figure~\ref{fig:synthetic_profiles}) illustrate the anomaly's morphological evolution with height. Key observations include:
\begin{itemize}
  \item Rapid decay of peak amplitude, particularly below 2\,m.
  \item Progressive increase in FWHM: from 1.5\,m at $h=0.2$\,m, to approximately 2.8\,m at $h=1.0$\,m, and nearly 4\,m at $h=5.0$\,m, representing more than a twofold broadening.
  \item Transformation from a sharp central maximum into a low contrast, gently rounded dome at higher altitudes.
\end{itemize}

\begin{figure}[!t]
\centering
\preprintgraphic[width=\columnwidth]{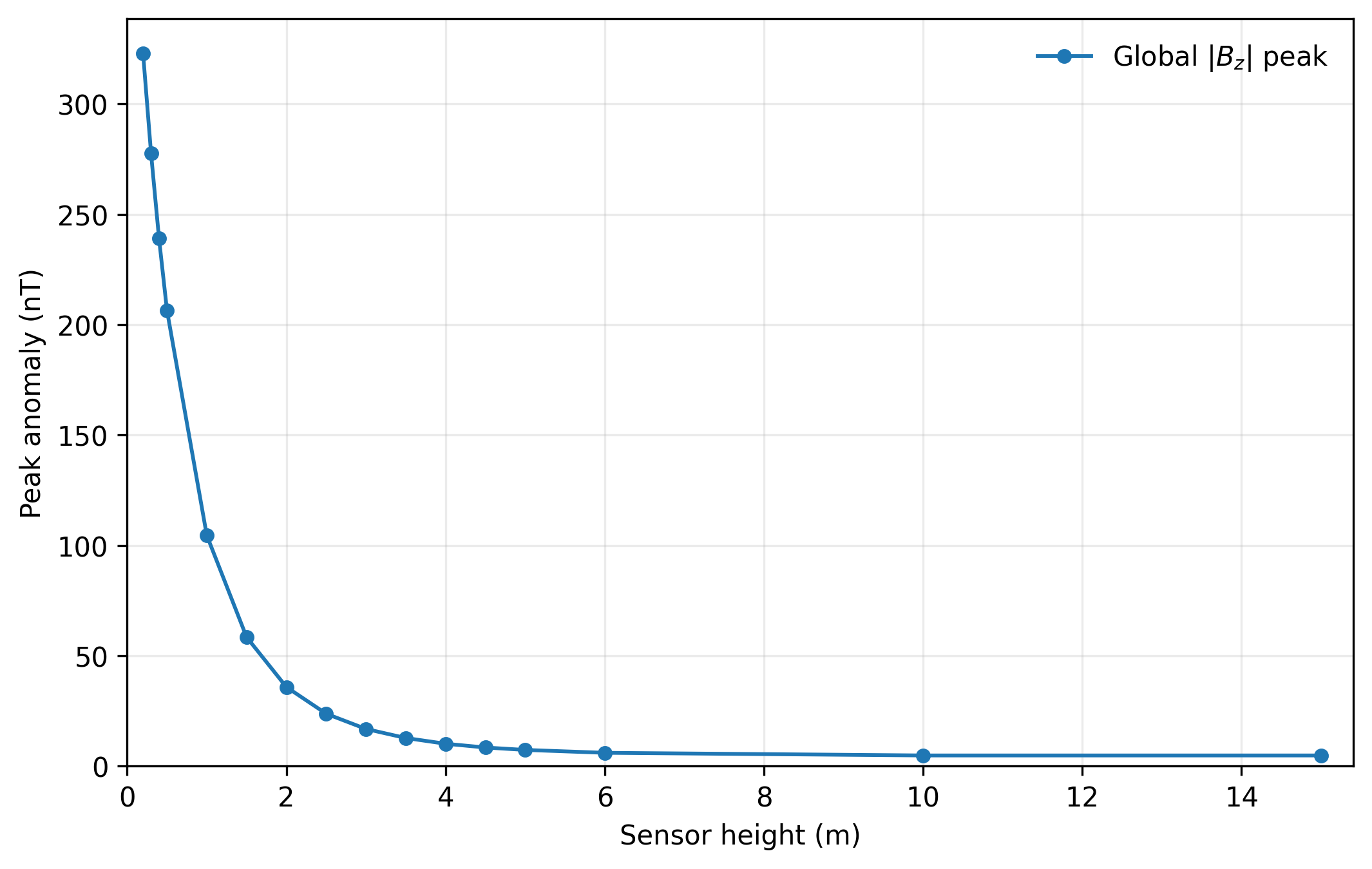}
\caption{Directly simulated global peak amplitude of the $|B_z|$ anomaly with increasing sensor height. No parametric decay model is imposed; the largest continuation heights approach a background-dominated level in the synthetic experiment.}
\label{fig:peak_decay}
\end{figure}

Figure~\ref{fig:peak_decay} shows the directly simulated peak amplitudes as a function of height without imposing a parametric decay law. The curve emphasizes the very rapid attenuation over the first few metres and the transition toward a background-dominated level at the greatest continuation heights.

\begin{figure}[!t]
\centering
\preprintgraphic[width=\columnwidth]{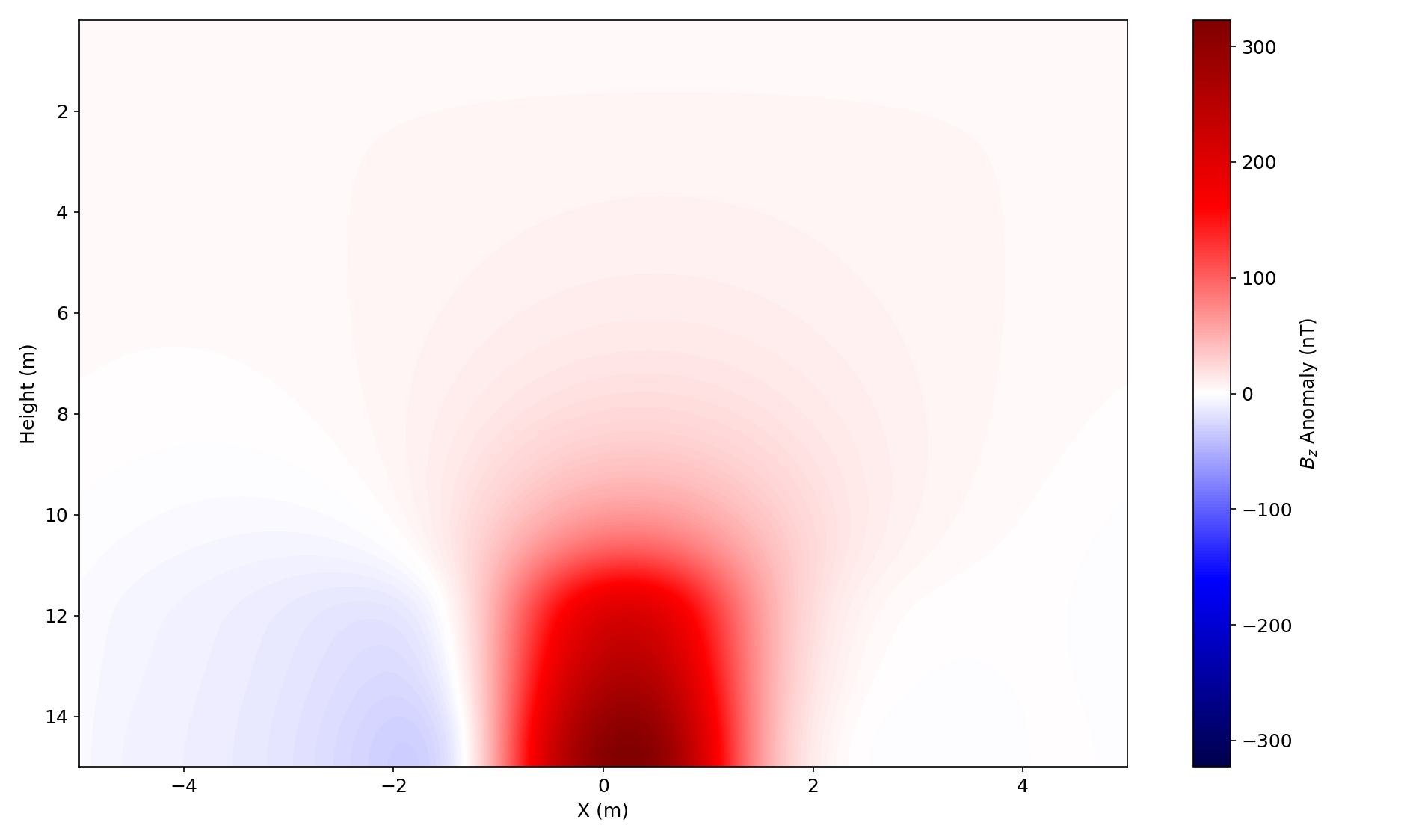}
\caption{Heatmap of the $B_z$ anomaly profile as a function of horizontal position and sensor height, illustrating both amplitude decay and lateral spreading during upward continuation.}
\label{fig:profile_heatmap}
\end{figure}

The heatmap (Figure~\ref{fig:profile_heatmap}) provides a comprehensive visualization of both the continuous amplitude decay and the lateral spreading with increasing height.

\subsection{Hybrid Continuation of Measured Anomaly}
The hybrid continuation experiment utilized real fluxgate gradiometer data from the Storbekken 1 site. This approach provides a more complex and realistic initial anomaly compared to the synthetic model.

\subsubsection{Plan-view Maps and Amplitude Decay}
\begin{figure*}[!t]
\centering
\preprintgraphic[width=0.96\textwidth]{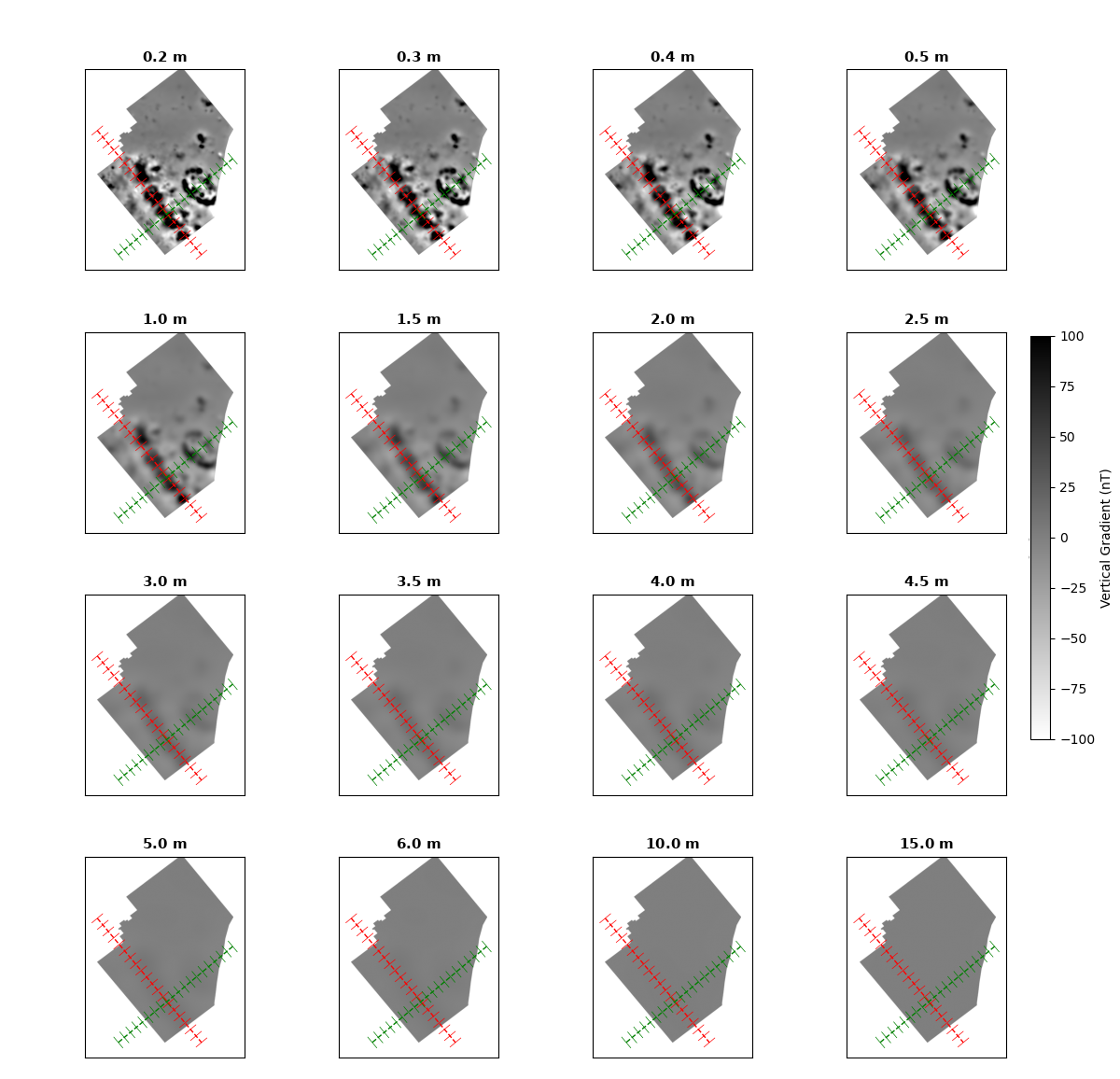}
\caption{Upward-continued vertical-gradient anomaly maps at selected sensor heights, illustrating the progressive reduction in amplitude and loss of spatial definition with elevation along Profiles~1 and 2.}
\label{fig:hybrid_maps_profiles}
\end{figure*}

At $h=0.2$\,m (as shown in Figure~\ref{fig:hybrid_maps_profiles}), the measured anomaly map displays multiple intense peaks and associated negative halos. Along Profile~1, the maximum anomaly reaches 519.9\,nT (100\%) and the minimum is $-141.8$\,nT (100\%). These strong positive and negative anomalies correspond to buried furnaces and slag deposits, exhibiting a spatially overlapping and irregular pattern distinct from the synthetic `bull's eye.'

Table~\ref{tab:hybrid_decay} presents the amplitude decay for both maximum and minimum anomalies along Profile~1 and Profile~2. The anomaly structure changes dramatically with increasing sensor height. At $h=0.5$\,m, the maximum and minimum have already decreased to 299.8\,nT (57.7\%) and $-75.8$\,nT (53.4\%) respectively, indicating a loss of over 40\% in only 30\,cm of elevation. By $h=1.0$\,m, the maximum is reduced to 141.2\,nT (27.2\%), and the minimum to $-36.8$\,nT (26.0\%). At $h=2.0$\,m (also visible in Figure~\ref{fig:hybrid_maps_profiles}), the formerly distinct multiplet of positive and negative lobes has largely merged into a single, broad positive high (45.9\,nT, 8.8\%) and a shallow negative low ($-13.4$\,nT, 9.5\%). By $h=5.0$\,m, only 7.5\,nT (1.4\%) of the initial peak persists, with the negative minimum at $-2.4$\,nT (1.7\%).

\begin{table*}[!t]
\centering
\caption{Maximum and minimum anomaly amplitudes for Profiles~1 and 2 in the hybrid experiment as a function of continuation height. Percentages are relative to the $h=0.2$\,m baseline for each profile.}
\label{tab:hybrid_decay}
\scriptsize
\setlength{\tabcolsep}{3.2pt}
\renewcommand{\arraystretch}{1.08}
\begin{tabular}{@{}r rr rr rr rr@{}}
\toprule
\multirow{2}{*}{Height (m)} &
\multicolumn{2}{c}{Profile~1 Max} &
\multicolumn{2}{c}{Profile~1 Min} &
\multicolumn{2}{c}{Profile~2 Max} &
\multicolumn{2}{c}{Profile~2 Min} \\
& Value (nT) & (\%) & Value (nT) & (\%) & Value (nT) & (\%) & Value (nT) & (\%) \\
\midrule
0.20 & 519.9 & 100.0 & $-141.8$ & 100.0 & 274.4 & 100.0 & $-145.5$ & 100.0 \\
0.30 & 428.6 & 82.4 & $-115.1$ & 81.2 & 227.0 & 82.7 & $-118.5$ & 81.4 \\
0.40 & 356.8 & 68.6 & $-93.5$ & 65.9 & 191.4 & 69.8 & $-97.2$ & 66.8 \\
0.50 & 299.8 & 57.7 & $-75.8$ & 53.4 & 163.9 & 59.7 & $-80.3$ & 55.2 \\
1.00 & 141.2 & 27.2 & $-36.8$ & 26.0 & 86.9 & 31.7 & $-33.2$ & 22.8 \\
1.50 & 76.7 & 14.7 & $-22.6$ & 15.9 & 52.9 & 19.3 & $-21.0$ & 14.5 \\
2.00 & 46.0 & 8.8 & $-13.4$ & 9.5 & 37.8 & 13.8 & $-14.5$ & 10.0 \\
2.50 & 29.7 & 5.7 & $-7.8$ & 5.5 & 29.7 & 10.8 & $-7.8$ & 5.5 \\
3.00 & 21.6 & 4.1 & $-6.0$ & 4.2 & 20.3 & 7.4 & $-7.3$ & 5.0 \\
3.50 & 16.1 & 3.1 & $-4.7$ & 3.3 & 15.3 & 5.6 & $-5.8$ & 4.0 \\
4.00 & 12.2 & 2.4 & $-3.7$ & 2.6 & 11.8 & 4.3 & $-4.7$ & 3.2 \\
4.50 & 9.5 & 1.8 & $-3.0$ & 2.1 & 9.3 & 3.4 & $-3.9$ & 2.6 \\
5.00 & 7.5 & 1.4 & $-2.4$ & 1.7 & 7.5 & 2.7 & $-3.2$ & 2.2 \\
6.00 & 5.0 & 1.0 & $-1.7$ & 1.2 & 5.0 & 1.8 & $-2.3$ & 1.6 \\
10.00 & 1.7 & 0.3 & $-0.5$ & 0.3 & 1.7 & 0.6 & $-0.8$ & 0.5 \\
15.00 & 0.8 & 0.2 & $-0.0$ & 0.0 & 0.8 & 0.3 & $-0.1$ & 0.1 \\
\bottomrule
\end{tabular}
\end{table*}

\subsubsection{Decay Curves and Morphological Evolution}
\begin{figure*}[!t]
\centering
\preprintgraphic[width=0.88\textwidth]{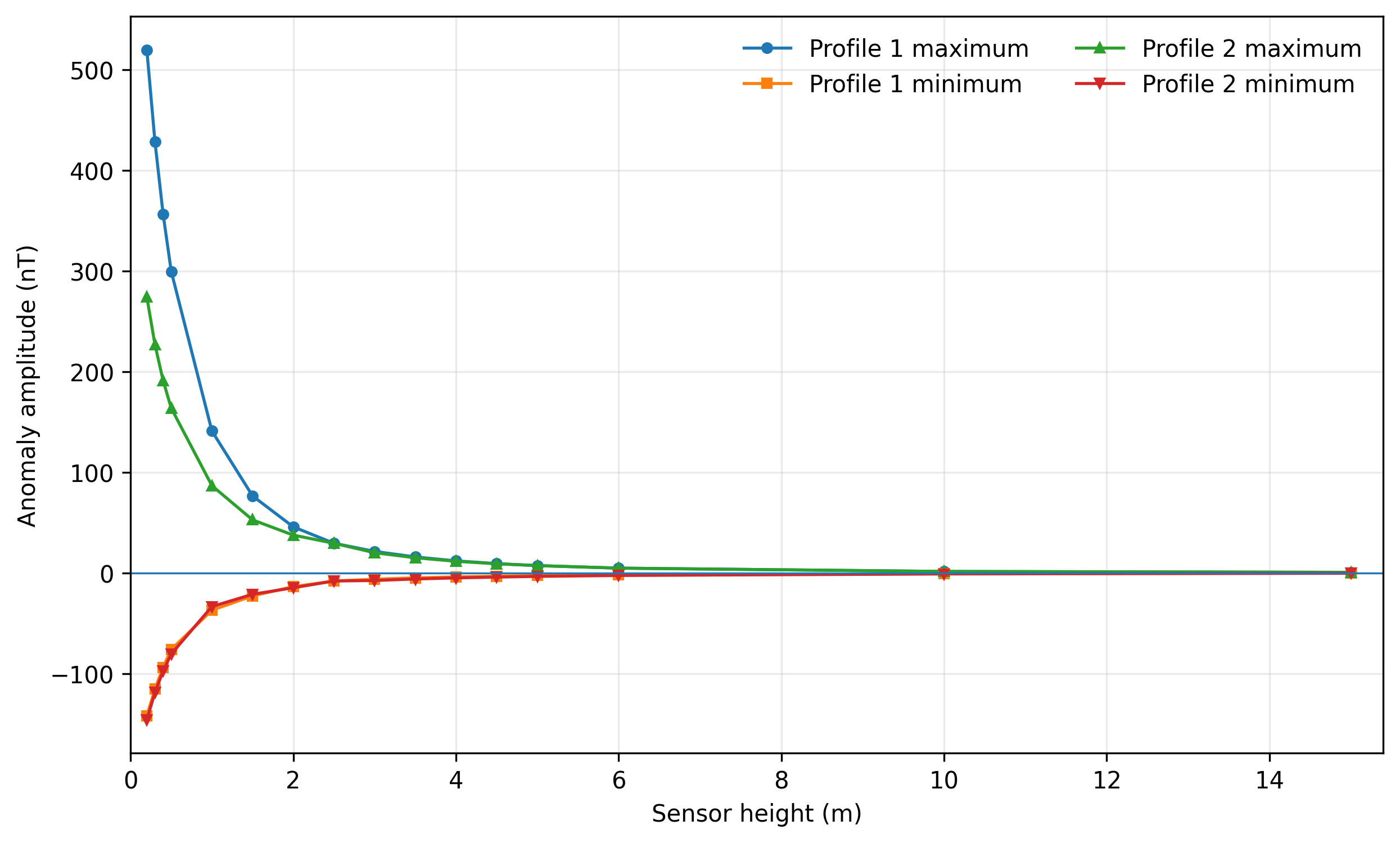}
\caption{Absolute anomaly amplitudes for Profiles~1 and 2 as a function of sensor height. Lines connect the directly continued minimum and maximum values.}
\label{fig:absolute_decay_fits}
\end{figure*}

\begin{figure*}[!t]
\centering
\preprintgraphic[width=0.88\textwidth]{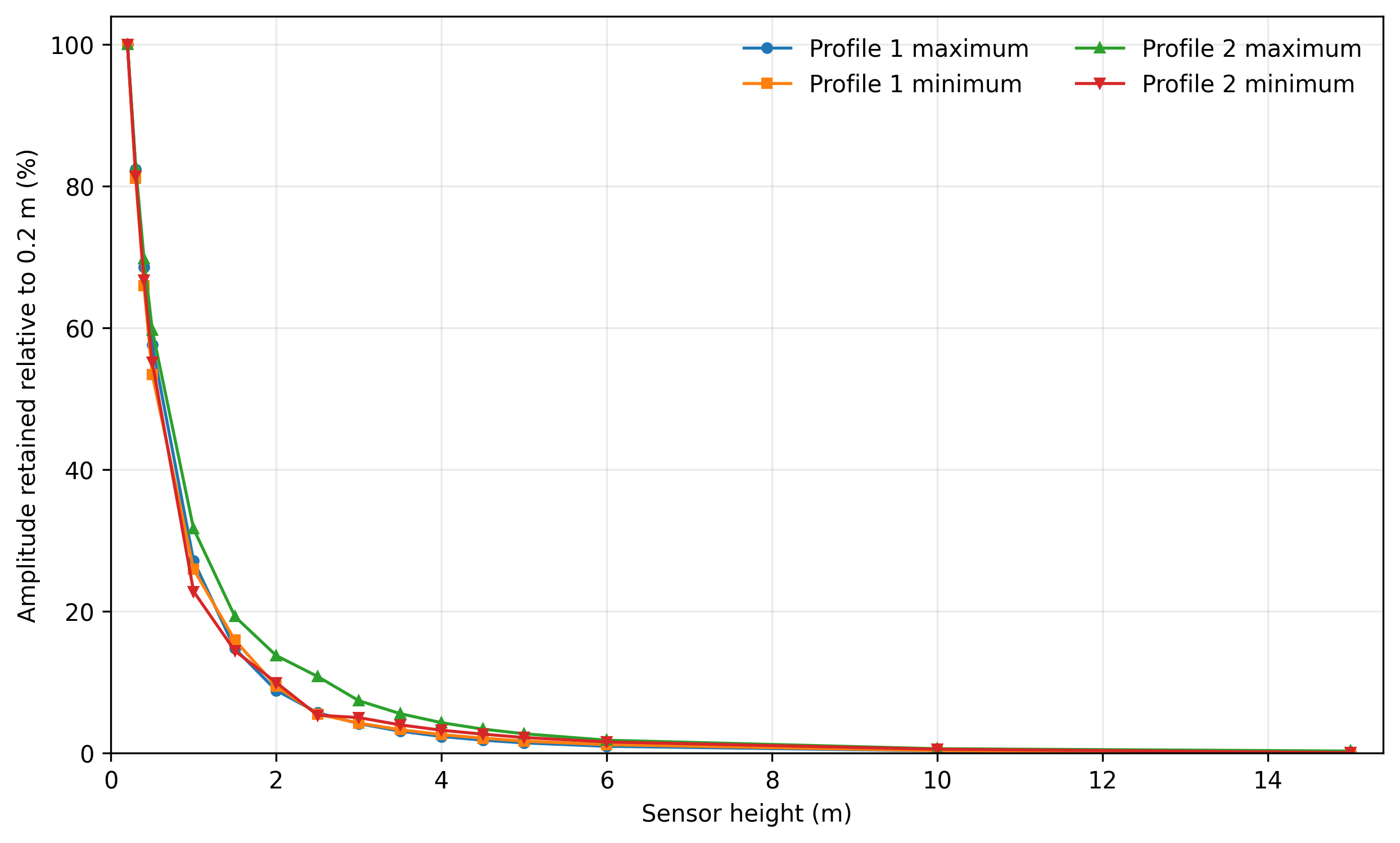}
\caption{Relative attenuation of the anomaly extrema for Profiles~1 and 2. Magnitudes are normalized to their respective values at $h=0.2$\,m.}
\label{fig:relative_decay_fits}
\end{figure*}

The four directly observed attenuation curves (Figures~\ref{fig:absolute_decay_fits} and \ref{fig:relative_decay_fits}) show the same rapid loss of amplitude with increasing height. At 2\,m, Profile~1 retains 8.8\% of its baseline positive maximum and 9.5\% of the magnitude of its baseline negative minimum, while Profile~2 retains 13.8\% and 10.0\%, respectively. These normalized values are used descriptively rather than being reduced to a fitted decay exponent.

The initial rate of amplitude decay is notably steep in the hybrid model: over 40\% of the signal is lost in just 30\,cm of elevation increase (from 0.2 to 0.5\,m), with an additional 30\% disappearing in the subsequent 50\,cm (0.5 to 1.0\,m). By $h=2.0$\,m, less than 10\% of the initial amplitude remains.

The plan-view morphology evolves rapidly with height (Figure~\ref{fig:hybrid_maps_profiles} and profile plots). At low altitude ($h=0.2$\,m), the map displays multiple sharp, high contrast peaks and steep gradients, characteristic of the clustered five buried furnaces and associated slag. As height increases to $h=1$--$2$\,m, these discrete features lose definition and merge into a single, broad, oval-shaped anomaly spanning 6--8\,m. Fine-scale dipolar halos and negative minima largely vanish by $h=2$\,m. At $h=4$--$5$\,m, the anomaly is reduced to a very faint, low contrast bulge, with all substructure lost and only approximately 1–2\% of the original amplitude persisting. Compared to the synthetic model, the real data exhibit stronger anisotropy, with anomaly broadening most pronounced along the NE–SW axis, consistent with the physical alignment of the furnaces. Lateral gradients and distinctions between sources are lost more rapidly in the hybrid model, resulting in a broad, indistinct high-altitude morphology.

\section{Discussion}
Our hybrid-continuation results, when viewed alongside the findings of \citet{Stele2023}, \citet{Stele2026}, and \citet{Schmidt2024}, provide a coherent picture of the primary physical and operational constraints affecting drone-borne magnetometry in archaeological contexts. These studies, together with our own and others such as \citet{Gavazzi2021} and \citet{accomando_drone_2024}, highlight how source--sensor distance, platform noise, acquisition geometry, and environmental conditions jointly control detectability and interpretability.

Recent field experiments by \citet{Stele2026} provide independent operational support for the altitude-dependent limitations identified here. Using a UAV-mounted multi-channel three-axis magnetometer system, they show that the UAV platform can act as both a source and an amplifier of magnetic noise, and that acquisition altitude, flight direction, velocity, and platform configuration materially affect data quality. Their recommendation to maintain the magnetometer array well below 1\,m AGL whenever operationally feasible is consistent with the rapid loss of anomaly amplitude and spatial definition in our synthetic and hybrid continuation experiments. The two studies address complementary components of the same problem: the present work quantifies source-distance attenuation and spatial smoothing, whereas \citet{Stele2026} demonstrates how platform- and acquisition-related noise further limits practical detectability. Accordingly, the simple sensor-noise realization used here should be interpreted as an idealized component of the observing system rather than a complete representation of UAV-system noise.

Firstly, all studies consistently underscore that drone-induced electromagnetic interference significantly impacts the magnetic data quality, particularly at higher frequencies (above a few hertz). \citet{Stele2023} comprehensively demonstrated strong noise peaks at specific frequencies (20\,Hz, 40–60\,Hz, and 82\,Hz) directly attributable to motor currents and rotor harmonics. These disturbances persist into the crucial low-frequency band relevant for archaeological features, necessitating strict mitigation measures such as 2\,Hz low-pass filtering and careful sensor placement beneath the airframe. \citet{accomando_drone_2024} also highlight noise issues due to magnetic interference from the mobile platform and oscillation of suspended sensors, which they addressed by suspending the magnetometer 3\,m below the drone and using low-pass filters. While our hybrid model incorporated Gaussian smoothing post-continuation to approximate such filtering, it is crucial to recognize that even under these idealized noise assumptions, furnace anomalies of only a few nanoteslas at higher altitudes closely approach realistic noise floors (0.3–0.5\,nT). This starkly reinforces the imperative of using ultra-low flight heights to maintain adequate signal to noise ratios and ensure detectability of anomalies that are inherently weak and rapidly decaying with altitude. \citet{Schmidt2024}, while primarily focusing on challenging terrain and OPMs, also implicitly acknowledge the need for low noise platforms to fully leverage sensor sensitivity, as environmental noise and drone-induced noise are major limiting factors for high resolution data.

Secondly, the performance of sensor resolution and drift in field conditions can markedly differ from laboratory specifications. \citet{Stele2023} observed variations of up to $\pm0.5$\,nT over minutes among their five R4 fluxgate sensors, noting that inner sensors exhibited greater drift than peripheral ones. This finding is critical when considering the typical resolution claims of magnetometers. In contrast, our modeled white noise $\sigma\approx0.009$\,nT represents an idealized intrinsic sensor-noise component rather than the full UAV-system noise; real surveys must contend with approximately $0.3$\,nT flicker noise and diurnal variations. Despite these practical challenges, both our synthetic and hybrid continuations as well as \citet{Schmidt2024} manhole tests, which specifically highlight the effectiveness of their miniature total field magnetometers (including an OPM) for detecting small anomalies at low heights confirm that well configured fluxgate systems, when properly filtered, retain the capability to detect archaeological anomalies down to a few nanoteslas. \citet{accomando_drone_2024} further demonstrate that drone-borne vertical gradient data, with proper noise reduction, compare very well with ground-based magnetic measurements. This suggests that while noise is a persistent challenge, effective survey strategies can still yield valuable data. \citet{vilhelmsen_dronetowed_2024} also present a new drone-towed electromagnetic induction and magnetic gradient sensor system for subsurface characterization, emphasizing robust data processing and stochastic inversion to enhance interpretation for archaeological prospecting, as demonstrated in the detection of an ancient flint mine. \citet{davis_stranded_2022} also demonstrated the capabilities of a MAGPI ML-4 atomic magnetometer mounted on a drone for identifying buried shipwrecks in challenging shallow coastal environments. Looking ahead, next generation magnetometer systems with substantially improved SNRs and drift stability will enable even more detailed surveys at higher altitudes, and our study’s quantified noise-level and flight-height thresholds offer clear design targets for those future experiments.

Thirdly, the morphological smoothing and amplitude decay inherent to upward continuation and increasing flight height is a universal and highly impactful phenomenon. This idealized experiment establishes an upper bound for magnetic anomaly preservation in realistic scenarios. Even for a strong, 323\,nT anomaly originating from an idealized kiln at 0.2\,m height, the signal is reduced to less than 10\,nT above 5\,m. This suggests that weaker archaeological features (e.g., pits, ditches, small ferrous debris with 10–50\,nT surface anomalies) would become effectively invisible at considerably lower altitudes. Furthermore, the observed morphological broadening implies that even strong anomalies may merge into indistinct features, complicating the interpretation of complex archaeological landscapes. These findings underscore the critical need for low altitude, high resolution survey strategies to successfully detect and resolve small scale archaeological features. \citet{Stele2023} vividly demonstrated that even at a relatively low altitude of 0.45\,m AGL, small pits (initially 20–50\,nT) begin to blur, effectively disappearing by 0.75\,m. \citet{Schmidt2024} further quantified this effect, documenting a significant fivefold footprint expansion of anomalies between 0.5\,m and 10\,m. Our hybrid maps precisely reproduce this behavior, showing how distinct furnace lobes at 0.2–0.5\,m rapidly coalesce into a single 6–8\,m-wide dome by 2–3\,m and expand beyond 10\,m by 4–5\,m. The more pronounced anisotropy and rapid loss of lateral gradients observed in our hybrid data, compared to our synthetic model, further underscore the complexity of real world multi source scenarios. This rapid loss of spatial detail supports sub-metre operation wherever the objective is to resolve individual features on the order of a few metres or less, while recognizing that the feasible altitude will depend on terrain, platform safety, and the magnetic contrast of the targets. In an exploratory survey one often benefits from the superposition of multiple subsurface sources into a single, broad anomaly at altitude. This behavior is directly analogous to discrete gas leaks that coalesce into a large concentration plume detectable from above before individual leaks can be pinpointed at close range \citep{pirk_inferring_2022}.  Similarly, spatially separated magnetic anomalies merge aloft into a “magnetic plume,” which serves as an initial beacon during high altitude passes. Once this composite feature is located, targeted low altitude flights can then “zoom in” to resolve and discriminate the individual sources with the spatial fidelity required for detailed archaeological interpretation. 

The hybrid continuation results have direct and significant implications for archaeological prospections using UAV-borne magnetometers. Even with the strong magnetic signature of the Storbekken 1 furnaces \citep{stamnes_magnetic_2019}, the observed rapid amplitude decay demonstrates that only a small fraction of the ground level anomaly persists at heights above 2--3\,m. Specifically, at a 2\,m sensor altitude, the dominant anomaly along both profiles is reduced to less than 14\% of its initial amplitude, and at 4\,m, only 2–4\% remains. These rates suggest that features with weaker initial signatures (e.g., smaller hearths, pits, or ditches, which often produce anomalies of 10–50\,nT at the surface) will become undetectable even more rapidly, potentially vanishing at sensor heights above just 1\,m. Morphologically, the upward continuation results reveal that distinct peaks from individual furnaces quickly merge into a single, broad anomaly as height increases. This smoothing out effect significantly reduces the ability to resolve closely spaced or subtle features, risking the loss of crucial archaeological information, such as identifying multiple adjacent furnaces, pits, or artefact clusters within complex sites. These results highlight the risk of "invisibility" for archaeological features in routine UAV surveys flown at standard altitudes (e.g., 3–5\,m). Even highly magnetic furnaces, let alone weakly magnetic targets, may fall below both sensor noise thresholds and the practical limit of visual interpretation. The rapid decay and morphological blending observed in the hybrid experiment provide a cautionary baseline for survey planning at other archaeological sites: if strong, well characterized kilns at Storbekken 1 can become marginal at a few meters altitude, then more subtle or complex sites demand an even more rigorous approach to low altitude magnetometry. Still, as these sites constitute of larger areas of enhanced magnetic response, the overall magnetic contrast is sufficient for site detection and delineation.

Finally, while not a primary focus of our study, \citet{Schmidt2024}  highlight the additional practical complexities introduced by challenging terrain (e.g., dense vegetation, steep slopes). Such environments inherently limit the ability to maintain consistent ultra low altitudes, thereby exacerbating the issues of amplitude decay and morphological smoothing, and posing significant obstacles for ground based surveys. Consequently, drone based methods become a necessary alternative, despite their inherent limitations. \citet{accomando_drone_2024} also emphasize that UAVs provide uniform coverage of large areas and access to very steep terrain, saving time and reducing risks. Similarly, \citet{vilhelmsen_dronetowed_2024} note the high time cost efficiency and flexibility of drones for surveying inaccessible areas, making them attractive or sometimes the only feasible option for geophysical measurements. From a survey design perspective, these findings underscore the necessity of operating UAV-borne sensors as close to the ground as safely possible,  ideally below 2\,m and preferably near 1\,m—to detect and spatially resolve small or subtle archaeological features. For sites with denser or more heterogeneous remains, or where regulatory or topographic constraints limit flight altitude, additional measures should be considered, such as denser line spacing, repeated surveys at lower altitudes, or employing higher sensitivity sensors (e.g., optically pumped magnetometers).

\section{Conclusions}

Through controlled synthetic modeling and hybrid upward continuation of real archaeological magnetometry data, we have quantitatively demonstrated that magnetic anomalies undergo rapid and systematic amplitude decay and morphological degradation with increasing sensor altitude—effects that impose strict operational constraints on drone-borne archaeological prospection.

\subsection{Key Quantitative Findings}

\begin{itemize}
    \item A strong synthetic anomaly (322.7 nT at 0.2 m) decays to < 3\% of initial amplitude above 5 m altitude (Table~\ref{tab:synthetic_decay}), while real furnace signatures (519.9 nT peak) lose > 40\% amplitude within the first 30 cm of elevation increase and < 10\% remains at 2 m (Table~\ref{tab:hybrid_decay}).
\item Morphological broadening is pronounced: anomaly FWHM expands > 2.5× between 0.2 m and 5 m altitude in the synthetic model, with distinct furnace lobes (separated by 5–6 m) coalescing into indistinct composite features by 2 m altitude in the hybrid experiment.
    
    \item Small or weak archaeological anomalies lose useful contrast especially rapidly with height; their practical detectability above approximately 1--2\,m will depend strongly on target amplitude, platform noise, environmental noise, and survey geometry.
\end{itemize}

\subsection{Primary Operational Implication}

For detailed characterization of small or weak archaeological features, sub-metre sensor heights should be targeted wherever operationally feasible. At sensor heights of approximately 2--3\,m, strong or spatially extensive archaeological complexes may remain detectable, but individual sources increasingly merge and lose diagnostic morphology. At still greater heights, the survey progressively shifts from feature characterization toward broad site-scale detection. These limits are not universal thresholds: practical performance also depends on target strength, source depth, platform noise, terrain, flight geometry, and processing strategy.

\subsection{Broader Context and Practical Implications}

Our findings align with and quantitatively extend recent literature \citep{Stele2023, Stele2026, Schmidt2024, accomando_drone_2024, vilhelmsen_dronetowed_2024}, collectively establishing that successful UAV magnetometry requires:

\begin{enumerate}
    \item \textbf{Ultra-low flight altitude} to maintain signal-to-noise ratios as rapid decay pushes anomalies toward detection thresholds.
    \item \textbf{Rigorous noise mitigation} (low-pass filtering, sensor suspension ≥ 2 m from motors, electromagnetic shielding) to address drone-induced interference.
    \item \textbf{Dense spatial sampling} (≤ 0.5 m line spacing) to adequately resolve broadened anomalies and prevent aliasing.
    \item \textbf{Realistic expectations} regarding minimum detectable feature size and magnetic moment, particularly in terrain-constrained scenarios where altitude cannot be minimized.
\end{enumerate}

While challenging terrain may necessitate UAV surveys at altitudes (2–3 m) that degrade small-feature detection, drones remain valuable for rapid reconnaissance and site discovery where broad magnetic contrast persists (e.g., smelting complexes, kiln clusters). The documented "magnetic plume" effect—wherein discrete sources merge into detectable composite anomalies at altitude—supports hierarchical survey strategies: high-altitude reconnaissance followed by targeted ultra-low-altitude characterization.

\subsection{Future Outlook}

Continued advances in high-sensitivity magnetometers (especially compact OPMs with sub-picotesla resolution), real-time platform compensation, multi-sensor gradiometry, and sophisticated inversion algorithms \citep{pirk_inferring_2022} will progressively extend the operational envelope of drone-borne magnetometry. However, the fundamental physics of potential-field attenuation (Eq.~\ref{eq:upward_continuation}) ensures that proximity to sources will remain the dominant factor in survey success. Our quantified decay curves (Tables~\ref{tab:synthetic_decay} and \ref{tab:hybrid_decay}) and altitude thresholds provide empirical benchmarks for evaluating future technological improvements and optimizing survey design across diverse archaeological contexts.

Importantly, our results demonstrate that drone-borne magnetometry is well-suited for site detection and delineation, but not for detailed intra-site feature characterization at typical UAV altitudes (> 2 m). This distinction is critical for setting appropriate expectations and designing survey strategies that leverage the strengths of UAV platforms while recognizing their fundamental physical limitations.

\FloatBarrier
\section*{Data Availability}
The synthetic datasets generated during this study, together with all numerical workflows used for magnetic-field simulation, upward continuation, noise modelling, attenuation analysis, and visualization, are fully reproducible using the openly available Python software released by the authors. The complete source code, including parameter files and example scripts for both synthetic and hybrid experiments, is publicly accessible via the MagSim repository (\url{https://github.com/alexandruhegyi/MagSim}). The field data from the Storbekken~1 site used in the hybrid analyses are derived from previously published surveys and are cited accordingly; no new raw field measurements were collected for this study.

\section*{Conflict of Interest}
The authors declare that they have no conflict of interest.

\FloatBarrier
\bibliographystyle{plainnat}
\bibliography{references}

\end{document}